\documentclass{article}
\usepackage{spconf}          
\usepackage{amsmath,amssymb}
\usepackage{graphicx}
\usepackage{booktabs}
\title{RadioLunaDiff: Estimation of Wireless Network Signal Strength in Lunar Terrain}
\usepackage{multirow}
\usepackage{multicol}
\usepackage{subcaption}
\usepackage{makecell}
\usepackage{enumitem}
\usepackage{url}
\usepackage{breakurl}
\usepackage{hyperref}

\name{Paolo Torrado$^{1}$, Anders Pearson$^{1}$, Jason Klein$^{2}$, 
Alexander Moscibroda$^{3}$ and Joshua Smith$^{1}$
\address{ \small
    $^{1}$University of Washington,
    $^{2}$Cornell University, 
    $^{3}$The Bear Creek School \\
    \small\texttt{\{patorrad, amp206\}@uw.edu, jak532@cornell.edu, alexmosci@outlook.com, jrs@cs.washington.edu}
}}

\begin{document}

\maketitle

\begin{abstract}
In this paper, we propose a novel physics-informed deep learning architecture for predicting radio maps over lunar terrain. Our approach integrates a physics-based lunar terrain generator, which produces realistic topography informed by publicly available NASA data, with a ray-tracing engine to create a high-fidelity dataset of radio propagation scenarios. Building on this dataset, we introduce a triplet-UNet architecture, consisting of two standard UNets and a diffusion network, to model complex propagation effects. Experimental results demonstrate that our method outperforms existing deep learning approaches on our terrain dataset across various metrics. The project website is available at: \href{https://radiolunadiff.github.io/}{https://radiolunadiff.github.io/}.

\end{abstract}


\section{Introduction}
The NASA proposed lunar exploration communication framework, LunaNet \cite{israel2020lunanet}, will require prediction of wireless network properties to facilitate reliable and robust communication. In a lunar environment, LunaNet is designed as a cooperative network infrastructure composed of multiple interoperable elements, where nodes communicate with neighbors and relay data at the appropriate link or network layers. Of particular importance are Radio Maps (RMs), which describe the spatial distribution of wireless channel attenuation for a given frequency, environment, and transmitter location.

For example, RMs can help mission planners create attenuation-aware trajectories or determine the shortest path to reconnection in cases of signal loss or degradation. Relaying rovers can also leverage RMs to maintain adequate signal quality when transmitting large data payloads, reducing energy consumption and communication latency.

\begin{figure}[!t]
  \centering
  \includegraphics[width=0.48\columnwidth]{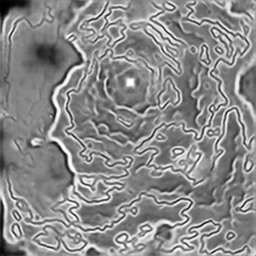}
  \hfill
  \includegraphics[width=0.48\columnwidth]{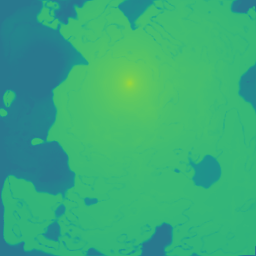}
  \caption{(a) Predicted $k^2$ probability map over lunar surface, (b) Corresponding model prediction of radio map at 5.8 GHz.}
  \label{fig:hero}
  \vspace{-1.45em} 
\end{figure}

Navigation is a key pillar of LunaNet, enabling the reliable fulfillment of safety, situational awareness, communication, and scientific objectives. Similar to GPS-RTK systems on Earth—where base stations provide correction data for centimeter-level accuracy—lunar rovers could use RMs to maintain robust data links with landers or beacons acting as reference nodes. In addition, RMs can support network reliability for critical services such as space weather monitoring and timely dissemination of solar eruption alerts. 
Such information is essential for protecting both equipment and astronauts.

Extensive prior research on wireless signal propagation has focused on predicting radio maps (RMs) in urban environments using deep neural networks. Leveraging simulation-based datasets, these studies have demonstrated that neural architectures can accurately estimate signal strength in complex scenes. More recently, researchers have employed physics-informed neural networks (PINNs) based on the Helmholtz equation to better model complex electromagnetic wave propagation, such as diffraction and interference, in dense environments.

In this work, we propose a novel triplet UNet PINN architecture that integrates the Helmholtz equation by modeling wave propagation over the terrain surface. Using a custom lunar terrain generator, we construct a simulation dataset and train a UNet to predict spatial variations in the square wave number, $k^2$, enabling the identification of electromagnetic field discontinuities induced by terrain geometry. This predicted $k^2$-map is then used as input to a second UNet chained to a diffusion model, both tasked with predicting the RM at either a low or high communication frequency, effectively incorporating terrain-induced propagation effects into the signal strength estimation process.
\begin{figure*}[t]
    \centering
    \includegraphics[width=0.95\textwidth, trim=0. 20 0 17, clip]{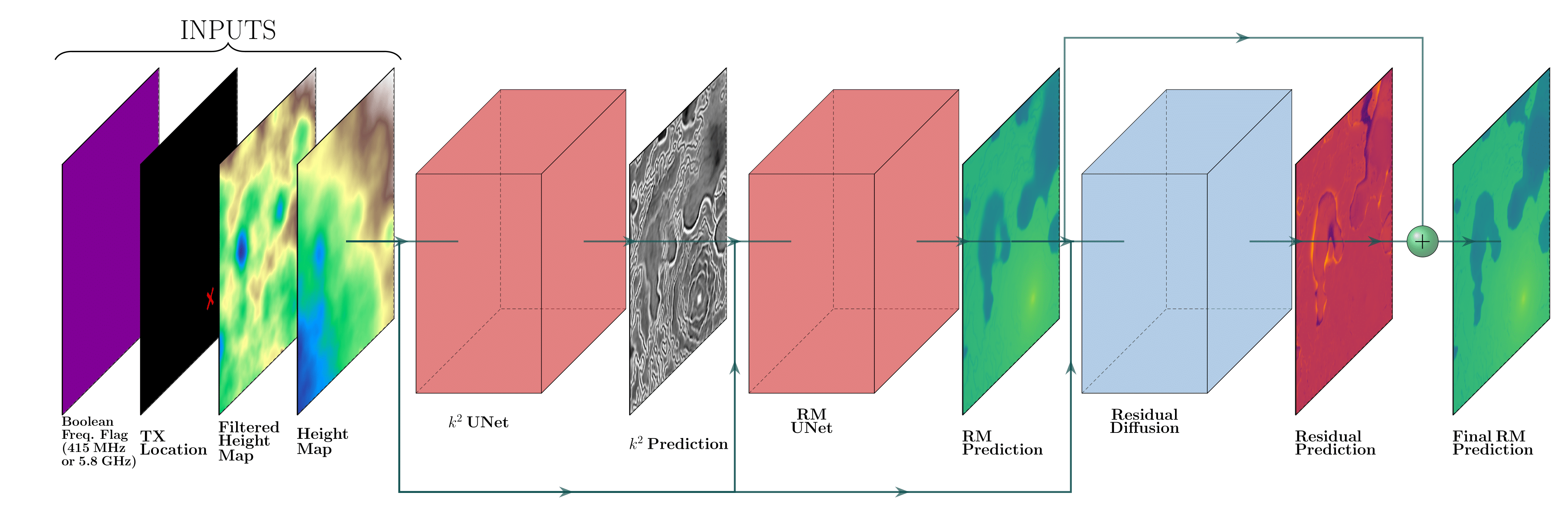}
    \caption{Proposed system architecture.}
    \small{Note: RMs shown in figure represent 5.8 GHz.}
    \label{fig:architecture}
    \vspace{-1.5em}
\end{figure*}
We make the following contributions:

\begin{itemize}[noitemsep, topsep=0pt]
    \item A lunar terrain generator that synthesizes realistic topography based on publicly available lunar elevation data.
    \item A novel physics-informed, diffusion-based model architecture for predicting radio maps over varying terrain at multiple frequencies.
\end{itemize}
\vspace{-1em}
\section{Related Works}
\vspace{-0.5em}
Traditional wireless propagation models, such as Rayleigh and TWDP Fading, estimate RMs in a statistical sense, sacrificing knowledge of the location of wireless hotspots and regions of attenuation for a simplified model that can only output a probability. While these models are effective in communications system design, applications such as localization and navigation require more detailed location information, leading to the development of methods that spatially sample and interpolate wireless channel attenuation \cite{rmInterp2021}. Unfortunately, this dense spatial sampling is often infeasible, leading recent research to focus on sampling-free methods like simulation.

Deterministic models such as ray tracers for radio propagation modeling, based on approximations of the Maxwell equations, have become increasingly precise and are widely used both commercially and in research \cite{sionna}. However, even with advances in efficiency and GPU acceleration, ray tracers remain computationally complex, leading researchers to explore deep learning for RM approximation.

One of the first successful approaches to deep learning-based RM estimation was \textit{RadioUNet}, which applied the UNet architecture to predict RMs using a map of building structures and the location of the transmitter \cite{radiounet2021}. Another notable adaptation, \textit{PMNet}, modified the traditional convolutional blocks of the UNet to incorporate atrous convolutions for improved feature extraction \cite{pmnet}. More recently, following trends in image generation, highly precise RM construction models have leveraged diffusion architecture \cite{radiodiff2024} \cite{jia2025rmdmradiomap} \cite{wang2025radiodiffk2}. 

Most existing deep learning approaches for radio map (RM) construction have been developed for planar environments, typically using simplified geometric descriptions of urban layouts represented by binary building maps. In contrast, this work focuses on non-planar RMs defined over complex lunar terrain, where elevation and surface irregularities play a dominant role in wave propagation. Prior efforts in this domain have been limited, often relying on simplified models such as the two-ray approximation for RM construction \cite{STAUDINGERlunarcomm}, which cannot fully capture the diffraction, refraction, shadowing, and scattering effects inherent to rugged lunar topography.
\vspace{-1em}
\section{Methods}
\vspace{-0.5em}
\subsection{Helmholtz Equation and Laplace-Beltrami Operator}
In this work we apply the Helmholtz Equation applied over a surface 
\begin{equation} \label{eq:helmholtz}
\Delta_{\mathcal{M}} E(\mathbf{x}) + k^2(\mathbf{x})E(\mathbf{x}) = -f(\mathbf{x)},
\end{equation}%

where $E(\mathbf{x})$ is the complex electric field, $k^2(\mathbf{x})$ is the wave number, $f(\mathbf{x})$ is the source term, and $\Delta_{\mathcal{M}}$ is the Laplace-Beltrami operator defined over the Riemannian manifold $\mathcal{M}$.
The Laplace-Beltrami operator is a generalization of the Laplace operator that works on arbitrary surfaces, as opposed to the Laplace operator which can only be used in Euclidean spaces \cite{jost2020riemannian} \cite{burman2018stablecutfiniteelement}.
\vspace{-1em}
\subsection{System Overview}
We propose a pipeline of three cascaded models—two UNets followed by a diffusion network (Fig.~\ref{fig:architecture})—to predict a radio map $I_{RM} \in \mathbb{R}^{H \times W}$. 
The input to the pipeline is denoted by $\mathbf{X}$ and consists of: 
(i) a height-map image $I_{HM} \in \mathbb{R}^{H \times W}$, 
(ii) a high-pass filtered height map $I_{FM} \in \mathbb{R}^{H \times W}$, 
(iii) a one-hot encoded image indicating the transmitter location $I_{Tx} \in \{0,1\}^{H \times W}$, and (iv) a boolean image flag $I_{Hz} \in \{0,1\}^{H \times W}$ which is all 0 to indicate a 415 MHz target frequency and all 1 for 5.8 GHz.


Next, we derive the square wave number map $k^2(\mathbf{x})$, represented in discrete form as $I_{KM} \in \mathbb{R}^{H \times W}$, by adapting RadioDiff-$k^2$ \cite{wang2025radiodiffk2} for surface data by rearranging ~\eqref{eq:helmholtz}. In practice, the source term $f(\mathbf{x})$ is determined by the transmitter location, which is not on the surface the electric field is defined on and therefore it is ignored. 

Let $I_E \in \mathbb{R}^{H \times W}$ denote the complex electric field on the terrain surface. 
Since the radio map $I_{RM}$ is a scalar field proportional to the squared field magnitude, $I_{RM}(i,j) \propto |I_{E}(i,j)|^2$, 
we approximate $k^2$ as:
\begin{equation} \label{eq:k2}
I_{KM}(i,j) = \frac{-\Delta_{\mathcal{M}}I_E(i, j)}{I_E(i, j)}.
\end{equation}
In practice, we utilize PyTorch and its auto-differentiable capabilities to derive the $I_{KM}$ \cite{paszke2017automatic}. The final step is to convert the KMs into binary values with pixels with values $<$ 0 being set to 1, and all non-negative pixels being set to 0.

Using $I_{HM}$, $I_{FM}$, $I_{Tx}$, and $I_{Hz}$ as inputs, the first UNet, $f_{\theta_1}$, estimates $\hat{I}_{KM}$ according to the formulation in~\eqref{eq:k2}. The second UNet, $f_{\theta_2}$, takes as input these three maps together with the predicted $\hat{I}_{KM}$ from the first UNet to estimate $\hat{I}_{RM}$. Finally, the third model, a Denoising Diffusion Probabilistic Model (DDPM) denoted $f_{\theta_3}$ estimates the residual map $\hat{r}$. The final prediction is reconstructed as:
\begin{equation}
\hat{I}_{\text{RM}}^{\text{final}} = \hat{I}_{\text{RM}, \theta_2} + \hat{r}_{\theta_3}.
\end{equation}%

\textbf{Stage 1: Physics-informed diffusion for wave number prediction.}
The first UNet, $f_{\theta_1}$, estimates the binary square wave number map $\hat{I}_{KM}$ 
from $\mathbf{X}$ and is trained with Binary Cross-Entropy (BCE) loss against the ground-truth ${I}_{KM}$, 
\begin{equation}
    \mathcal{L}_1 = -\frac{1}{HW} \sum_{i=1}^H \sum_{j=1}^W \Big[ 
    BCE({I}_{KM, ij}, \hat{I}_{KM, ij}) \Big].
\end{equation}
which compels the model to output a probability map $\hat{I}_{KM}$. Each pixel in this map represents the model's confidence that a $k^2$ discontinuity is present. While this map could be binarized using a 0.5 threshold, we retain the continuous probabilities as a "soft" input for the subsequent stage, thereby preserving the model's uncertainty.

\textbf{Stage 2: Initial Radio Map Prediction.}
The second UNet, $f_{\theta_2}$, produces an initial estimate of the radio map $\hat{I}_{RM, \theta_2}$ from the inputs $\mathbf{X}$ and the predicted square wave number map $\hat{I}_{KM, \theta_1}$. 
The model is trained by minimizing the Mean Squared Error (MSE) loss:
\begin{equation}
    \mathcal{L}_2 = \frac{1}{HW} \sum_{i=1}^H \sum_{j=1}^W 
    \big(I_{RM,ij} - f_{\theta_2}(\mathbf{X}, \hat{I}_{KM,\theta_1})_{ij}\big)^2.
\end{equation}

\textbf{Stage 3: Diffusion for Residual Refinement.}
In the final state, we utilize the architecture DDPM as proposed by Ho et al. \cite{ho2020ddpm} and used in \cite{radiodiff2024} to act as a refinement model. The model, $f_{\theta_3}$, which refines the initial radio map by predicting the residual 
$r = I_{RM} - \hat{I}_{RM,\theta_2}$. 
Its UNet core $U_{\theta_3}$ takes a noisy residual $r_t$ at timestep $t$ together with conditioning inputs 
$c=(\mathbf{X}, \hat{I}_{KM,\theta_1}, \hat{I}_{RM,\theta_2})$, and predicts the velocity $v_{\text{pred}}$. 
The model is trained with a hybrid objective that combines velocity prediction \cite{salimans2022progressivedistillationfastsampling} and residual reconstruction:
\begin{equation}
    \mathcal{L}_3 = \mathcal{L}_v + \lambda \mathcal{L}_{\text{recon}},
\end{equation}
where the MSE between the true velocity $v$ and the network's prediction, $v_{pred}$, is
\begin{equation}
    \mathcal{L}_v = \mathbb{E}_{r,\epsilon,t,c}\big[ \| v - U_{\theta_3}(r_t,t,c)\|^2 \big],
\end{equation}
and the MSE between the true residual $r$ and the predicted residual , $\hat{r}_{\theta_3}$, is
\begin{equation}
    \mathcal{L}_{\text{recon}} = \mathbb{E}_{r,\epsilon,t,c}\big[ \|r(\mathbf{x}) - \hat{r}_{\theta_3}\|^2 \big].
\end{equation}
The predicted residual is obtained from the velocity estimate as
\begin{equation}
    \hat{r}_{\theta_3} = \sqrt{\bar{\alpha}_t}\,r_t - \sqrt{1-\bar{\alpha}_t}\,v_{\text{pred}}.
\end{equation}

\textbf{Overall objective.}
The final loss is:
\begin{equation}
\mathcal{L}_{\text{total}} = 
\underbrace{\mathcal{L}_{1}}_{\text{Stage 1: $k^2$ map}}
+ \underbrace{\mathcal{L}_{2}}_{\text{Stage 2: radio map}} 
+ \underbrace{\mathcal{L}_3}_{\text{Stage 3: residual}}.
\end{equation}%

\textbf{UNet Architecture, Training and Inference.} 
$f_{\theta_1}$ is implemented using a custom UNet, $f_{\theta_2}$ is implemented by modifying the architecture from \textit{PMNet} \cite{pmnet}, and $f_{\theta_3}$ is implemented via architecture provided by the Hugging Face Diffusers python library \cite{von-platen-etal-2022-diffusers}.
Each stage is trained separately, with gradients from $\mathcal{L}_{2}$ blocked from flowing into $\theta_1$ by the stop-gradient on $\hat{I}_{KM}$, and gradients from $\mathcal{L}_{3}$ blocked from flowing into $\theta_2$ and $\theta_1$ by the stop-gradient on $\hat{I}_{RM}$.
At inference, we first generate $\hat{I}_{KM}$ using $f_{\theta_1}$ and then condition $f_{\theta_2}$ on $\hat{I}_{KM}$ to produce $\hat{I}_{RM}$, and finally condition $f_{\theta_3}$ on $\hat{I}_{KM}$ and $\hat{I}_{RM}$ to produce $\hat{r}$, finally reconstructing $\hat{I}_{\text{RM}}^{\text{Final}}$.
\vspace{-0.5em}
\section{Experimental Setup and Datasets}
\vspace{-0.5em}
\subsection{Lunar Terrain Height Map Generation}
Lunar terrain is dominated by impact craters and the geomorphic consequences of cratering. We generate synthetic lunar terrains following Appendix~C of Cai and Fa \cite{cai2020moon} with numerical details following \cite{press2007numerical} \cite{neukum2001cratering} \cite{fassett2014degradation}. Specifically, we alternate between (i) simulating a subset of cratering events corresponding to a fraction of the target surface age and (ii) applying diffusion over that same interval. Iterating these two steps until the total age is reached yields complex terrains containing both fresh, sharply defined craters and older, highly degraded craters. 

\begin{figure*}[t]
    \centering
    
    \begin{subfigure}[b]{0.16\textwidth}
        \includegraphics[width=\textwidth]{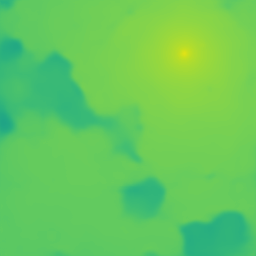}
        \caption{RadioUNet}
        \label{fig:radiounet}
    \end{subfigure}%
    \hfill
    \begin{subfigure}[b]{0.16\textwidth}
        \includegraphics[width=\textwidth]{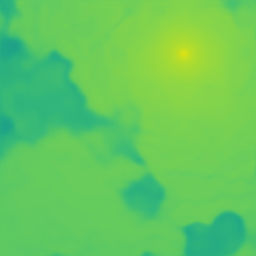}
        \caption{PMNet}
        \label{fig:pmnet}
    \end{subfigure}%
    \hfill
    \begin{subfigure}[b]{0.16\textwidth}
        \includegraphics[width=\textwidth]{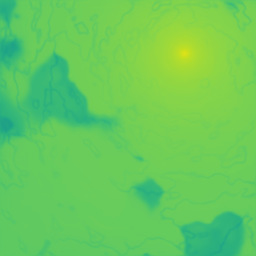}
        \caption{w/o Diff}
        \label{fig:no_diff}
    \end{subfigure}%
    \hfill
    \begin{subfigure}[b]{0.16\textwidth}
        \includegraphics[width=\textwidth]{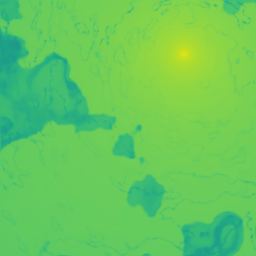}
        \caption{RM Diff}
        \label{fig:rm_diff}
    \end{subfigure}%
    \hfill
    \begin{subfigure}[b]{0.16\textwidth}
        \includegraphics[width=\textwidth]{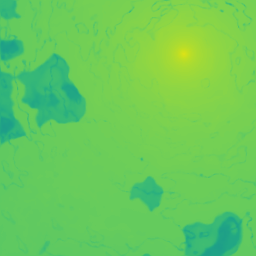}
        \caption{Ours}
        \label{fig:ours_final}
    \end{subfigure}%
    \hfill 
    \begin{subfigure}[b]{0.16\textwidth}
        \includegraphics[width=\textwidth]{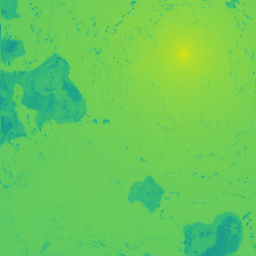}
        \caption{Ground Truth}
        \label{fig:gt}
    \end{subfigure}

    \caption{\textbf{Qualitative comparison of generated radiomaps at 415 MHz.} }
    \label{fig:qualitative}
    \vspace{-1em} 
\end{figure*}
\vspace{-1em}
\subsection{Sionna Simulation Dataset}
We built our dataset using Sionna-RT, an open source raytracer for radio propagation modeling \cite{sionna}. Based on LunaNet specifications we choose frequencies 5.8GHz and 415MHz \cite{Israel2023LunaNet}. Ray-tracing methods are inherently limited by the finite number of discretized rays, which often results in radio maps with null pixels. To mitigate this, we randomize the seed and recompute each map instance multiple times.
Bilinear interpolation was used to complete 415~MHz maps, which exhibited scattered data gaps, while a static fill was applied to 5.8~GHz maps, which contained large contiguous voids and few scattered points.

The electromagnetic properties of lunar regolith—relative permittivity $\varepsilon'_{\text{reg}}$ (real part) and electrical conductivity $\sigma$—are taken from NASA specifications \cite{sionna}.


\section{Evaluation and Results}

\subsection{Metrics}
We evaluate pixel-wise performance using the root mean squared error (RMSE), normalized mean squared error (NMSE), structural similarity index (SSIM) and peak signal-to-noise ratio (PSNR), as used in \cite{firstrmprediction2023} \cite{radiodiff2024}  \cite{wang2025radiodiffk2}.
\subsection{Quantitative Results}
To evaluate our model, we compared it against neural network–based RM prediction methods that could be easily retrained with our dataset. 
Specifically, we selected PMNet \cite{pmnet} and RadioUNet \cite{radiounet2021}. 
PMNet achieved the best performance in the First Pathloss Radio Map Prediction Challenge \cite{firstrmprediction2023}, while RadioUNet was among the first and most effective NN-based RM prediction models. 
More recent architectures such as RMDM \cite{jia2025rmdmradiomap}, RadioDiff \cite{radiodiff2024}, and RadioDiff-$k^2$ \cite{wang2025radiodiffk2} leverage diffusion models but require substantial modification for our terrain setting or do not yet have publicly available code. 

We evaluated all models at 415~MHz, 5.8~GHz and on combined results across both bands. 
Table~\ref{tab:quantitative} reports the quantitative metrics, where our method consistently outperforms the baselines across all metrics, with the greatest difference in SSIM and PSNR.
In particular, our approach yields lower error and better structural fidelity, demonstrating its superiority over PMNet and RadioUNet. We performed an ablation study by either removing the final residual diffusion step (w/o Diff) or replacing the residual target with the RM itself (RM Diff). As shown in Table~\ref{tab:quantitative}, both modifications degraded performance compared to our full model.

A qualitative comparison further highlights the advantages of our model. 
As shown in Fig.~\ref{fig:qualitative}, our predictions preserve fine geometric details and accurately localize electromagnetic singularities, especially in low-connectivity regions such as craters and terrain shadowed by hills. 
In contrast, the baseline methods tend to blur sharp features in the radio map or misplace singularities, leading to degraded interpretability and less reliable coverage prediction.

\begin{table}[h!]
    \centering
    \caption{\textbf{Quantitative Comparison of Models}.}
    \captionsetup{skip=0pt} 
    \small{Note: Results in \textbf{\underline{bold-underlined}} and \underline{underlined} represent the best and second best scores, respectively.}
    \setlength{\tabcolsep}{5pt}
   
    \begin{tabular}{@{}c|ccccc@{}}
    \toprule
    &  \multicolumn{2}{c}{Comparative} & \multicolumn{2}{c}{Ablation} & \\ \midrule
    Methods & \makecell{Radio-\\UNet} & PMNet & \makecell{w/o\\Diff} & \makecell{RM\\Diff} & Ours   \\ \midrule
    415 MHz & & & & & \\ \midrule
    NMSE & .001060 & .000869 & \underline{.000820} & .000898 &  \textbf{\underline{.000795 }}\\
    RMSE & .0243 &  .0220 &  \underline{.0214} & .0224  & \textbf{\underline{.0211}} \\
    SSIM $\uparrow$ & .8810 & .8867 & \underline{0.9035} & 0.8975 &  \textbf{\underline{.9069}} \\
    PSNR $\uparrow$ & 33.54 & 34.26 & \underline{34.77} & 34.53 & \textbf{\underline{35.40}} \\ \midrule
    5.8 GHz & & & & & \\ \midrule
    NMSE & .00339 & \underline{.00258} & .00264 & .00270 &  \textbf{\underline{.00248}} \\
    RMSE &  .0354 & \underline{.0309} & .0312 & .0315 &  \textbf{\underline{.0302}}\\
    SSIM $\uparrow$ & .8826 & 0.8898 & .9009 & \underline{.9014} & \textbf{\underline{0.9067}}\\
    PSNR $\uparrow$ & 30.35 & 31.39 & 31.64 & \underline{32.10} & \textbf{\underline{32.65}} \\ \midrule
    Both & & & & & \\ \midrule
    NMSE &  .00223 &  \underline{.00172} & .00173 & .00180& \textbf{\underline{.00164}} \\
    RMSE & .0299 &  .0265 & \underline{.0263} & .0270 & \textbf{\underline{.0257}}\\
    SSIM $\uparrow$ & .8818 & .8883 & \underline{.9022} & .8995 & \textbf{\underline{.9068}}\\
    PSNR $\uparrow$ & 31.95 & 32.83 & 33.20 & \underline{33.32} & \textbf{\underline{34.03}} \\ \bottomrule
    \end{tabular}
    \label{tab:quantitative}
    \vspace{-1.5em} 
\end{table}

\section{Conclusion}\vspace{-0.2em}
In this work, we presented a physics-informed deep learning framework for predicting radio maps over complex lunar terrain. 
Our approach integrates a synthetic lunar terrain generator with a ray-tracing pipeline to construct training datasets, and introduces a physics based triplet-UNet architecture composed of two UNets and a diffusion model. 
Experimental results on simulated lunar data demonstrate that our method improves accuracy over baseline deep learning models. Furthermore, we carried out ablation studies to show the need for the $k^2$ map. The ability to generate reliable radio maps over lunar terrain supports future communication-aware mission planning for the LunaNet framework.

\section{ACKNOWLEDGMENT}
This work was supported by NASA Grant 80NSSC22K0258 to the University of Washington, with a subcontract to Astrobotic Technology. Computations were performed on Hyak, the University of Washington’s high-performance cluster funded by the student technology fee.


{\footnotesize
\bibliographystyle{IEEEtran}
\bibliography{root}
}

\end{document}